\documentclass[epj-spec]{svjour}
\usepackage{graphics}

\usepackage{ulem}
\usepackage{xcolor}

\begin{document}

\title{Unconventional Quantum Phases in Lattice Bosonic Mixtures}
\author{P. Buonsante\inst{1} \and S.M. Giampaolo\inst{2,3} \and F.
Illuminati \inst{2,3,4} \and V. Penna\inst{1} \and A. Vezzani\inst{5,6}}
\institute{C.N.I.S.M. and Dipartimento di Fisica, Politecnico di
Torino, C.so Duca degli Abruzzi 24, I-10129 Torino, Italy. \and
Dipartimento di Matematica e Informatica, Universit\`a degli Studi
di Salerno, Via Ponte don Melillo, I-84084 Fisciano (SA), Italy. \and
CNR-INFM Coherentia, Napoli, Italy, CNISM Unit\`a di Salerno,
 and INFN Sezione di Napoli, Gruppo collegato di Salerno, Baronissi (SA), Italy.
 \and Institute for Scientific Interchange, Viale Settimio Severo 65, I-10133 Torino, Italy. \and
CNR-INFM, S3, National Research Center, via Campi 213/a, I-41100 Modena.\and
Dipartimento di Fisica, Universit\`a degli Studi di
Parma, V.le G.P. Usberti n.7/A, I-43100 Parma, Italy }

\abstract{We discuss strongly interacting one-dimensional boson-boson mixtures and study their quantum phases as the interspecies repulsion is increased.
In particular, we analyze the low-energy {\it quantum emulsion} metastable states occurring
at large values of the interspecies interaction, which are expected to prevent the system
from reaching its true ground state. A significant decrease in the visibility of the atomic
clouds is predicted as well, and related to the trapping of the system in the spontaneously
disordered quantum emulsion states. }

\maketitle

\section{Introduction} \label{sec.intro}

In recent years the study of systems of ultracold atoms loaded into
optical lattices has drawn much attention because of their value as
experimental probes for the investigation of models of condensed matter
physics exhibiting key properties of complex many-body systems.
One of the most relevant examples in this sense is certainly the
Bose-Hubbard model, originally introduced by Haldane
\cite{Haldane_PLA_80_280} as a simple yet interesting variant of the
better known fermionic Hubbard model, and later investigated at length
and proposed for the description of superfluid $^4$He trapped in porous
media in a seminal paper by M. P. A. Fisher et al. \cite{Fisher_PRB_40_546}.
The observation of the distinctive superfluid-Mott insulator
quantum phase transition in a system of ultracold rubidium atoms
\cite{Greiner_Nature_415_39} brilliantly demonstrated the prediction
that, under suitable conditions, ultracold bosonic atoms trapped in
optical lattices could provide a physical realization of the
Bose-Hubbard model \cite{Jaksch_PRL_81_3108}.
The impressive control in atom cooling and trapping soon allowed the
realization of more general Bose-Hubbard Hamiltonians, including, e.g.,
disordered local potentials \cite{Lye_PRL_95_070401,Clement_PRL_95_170409}.

The quest for  novel and unconventional quantum phases
has recently pushed the interest towards even more general variants of
the Hubbard model, involving mixtures of particles obeying either the same
or different statistics. Beyond their theoretical appeal, these systems are
relevant to interesting applications such as implementation of
disordered systems \cite{Gavish_PRL_88_170406,Ospelkaus_PRL_96_180403},
association of dipolar molecules \cite{Damski_PRL_90_110401},
schemes for quantum computation \cite{Daley_PRA_69_022306} and realization
of quantum spin chains and arrays \cite{Kuklov_PRL_90_100401,Duan_PRL_91_090402}.

So far, most of the experiments on lattice-atom mixtures have involved
atoms obeying different statistics. Several experimental realizations of lattice
Bose-Fermi mixtures have been engineered
\cite{Ospelkaus_PRL_96_180403,Ferlaino_JPhysIV_116_253,Gunter_PRL_96_180402}, whereas,
to the best of our knowledge, the first experiment on mixtures involving two different bosonic
species has been carried out very recently at LENS, in Florence \cite{Catani_PRA_77_011603}.

At the theoretical level, various investigations have been carried out on lattice Bose-Fermi 
\cite{Albus_PRA_68_023606,Adhikari_PRA_70_043617,Illuminati_PRL_93_090406,Cramer_PRL_93_190405,Roth_PRA_68_023604,Roth_PRA_69_021601}
as well as 
Bose-Bose mixtures
\cite{Altman_NJP_5_113,Chen_PRA_67_013606,Alon_PRL_97_230403,Mathey_PRB_75_144510,Roscilde_PRL_98_190402,Buonsante_PRL_100_240402}.
This wealth of work demonstrates that a comprehensive and systematic study of the phases
of atomic lattice mixtures requires a significant analytic and synthetic effort.
Indeed, one should be aware that even in the ideal case of a homogeneous lattice, the
Hamiltonian describing a bosonic mixture contains five parameters, namely the
two intra-species interaction strengths, the inter-species interaction strength
and one hopping amplitude for each species.
Moreover, the possible phases depend on many physical parameters such as, e.g.,
the populations of the two atomic species and their commensurability with the lattice size.

For the above reasons, 
we confine the present analysis to a rather limited region of the phase diagram of a boson-boson mixture, which however encompasses a fair wealth of 
quantum phases effects. Specifically, we consider
two different bosonic species loaded  in a 1-D homogeneous lattice,
and vary only the interaction among them. 

We assume that one of the species is soft-core and at unitary filling, whereas the other is hard-core and at filling $2/5$. Our choice is similar to
that considered in Refs. \cite{Pollet_PRL_96_190402,Hebert_PRA_76_043619}, where the
ground state of a Bose-Fermi mixture on a 1-D lattice is analyzed by means of quantum
Monte Carlo simulations.  
Recall indeed that on 1D lattices hard-core bosons and spinless fermions 
share many features.
While we find a qualitatively similar phase diagram, our analysis
differs from
that in Ref.~\cite{Hebert_PRA_76_043619} at least in two respects. Firstly, 
we adopt a less quantive yet more computationally affordable approach. 

More interestingly, we focus on the configurations that emerge at strong
values of interspecies interactions. We find that for sufficiently large
interspecies interactions the system supports a large number of low energy
metastable {\it quantum emulsion} states exhibiting glassy features
despite the complete absence of any source of disorder in the Hamiltonian parameters.
Such a complex energy landscape for lattice bosonic mixtures
was pointed out in Ref. \cite{Roscilde_PRL_98_190402} and further discussed
in Ref. \cite{Buonsante_PRL_100_240402} in the presence of the inhomogeneous
local potential typical of realistic setups. Interesting  related results  are
presented in Ref. \cite{Alon_PRL_97_230403} based on a multi-orbital
Gross-Pitaevskii approach, while a phase diagram for metastable states
is derived in Ref. \cite{Menotti_PRL_98_235301} for dipolar bosons.
We carry the analysis of quantum emulsions further on,
discussing the arrangement and features of the droplets,
as well as the interference pattern obtained after  free-expansion of the atomic cloud.

The plan of the paper is as follows: In Sec. \ref{sec.mf} we
recall the Hamiltonian of the system as well as the mean-field approach
we adopt.
After the introduction of the quantities characterizing the quantum phases
occurring in the system, our results are presented in
Sec. \ref{sec.rs}. In particular, we focus on the behavior of all the aforementioned  quantities as functions  of the interspecies interaction.
As we mention, particular attention is devoted to the quantum
emulsion states characterizing the phase-separated regimes.
Interestingly we evidence a phase where the soft-core bosons behave as a
{\it disordered superfluid} 
 featuring a global phase coherence but a spatially inhomogeneous density. 
We conclude with a brief summary of our results.

\section{The System} \label{sec.mf}

The arguments of  Ref. \cite{Jaksch_PRL_81_3108} can be generalized
to a lattice loaded with atoms of two different bosonic species.
Under suitable conditions, the system is described by the two-flavor
Bose-Hubbard Hamiltonian
\begin{equation}
\label{Hamiltonian} H = U_{1\,2} \sum_i n_{1,i}\, n_{2,i}+ \sum_{f}
\sum_i \left[\frac{U_f}{2} n_{f,i} (n_{f,i}-1)  - J_f \left(
a_{f,i}^\dag a_{f,i+1} + h.c. \right)\right] \; ,
\end{equation}
where the lattice boson operators $a_{f,i}^\dag$, $a_{f,i}$, and
$n_{f,i} = a_{f,i}^\dag \,a_{f,i}$, create, destroy and count atoms
of the flavor, or species, $f$ at site $i$. The parameters $U_f$ and
$U_{1\,2}$ quantify the repulsive interaction between atoms of the
same or different species, respectively (henceforth intra-
and inter- species repulsion). The possibly different hopping
amplitude of the two species is quantified by the parameters $J_f$.
Two-species Bose-Hubbard Hamiltonians similar to the one in Eq.
(\ref{Hamiltonian}) have been considered previously in several
works, possibly referring to different internal states of the same
bosonic species \cite{Chen_PRA_67_013606} to spin-1
\cite{Krutitsky_PRA_70_063610,Krutitsky_PRA_71_033623,Yamashita_PRA_76_023606}
or dipolar bosons
\cite{Damski_PRL_90_110401,Arguelles_PRA_75_053613}.

Since we are in a strong interaction regime, we assume that the
state of the system $|\Psi\rangle$ has a Gutzwiller factorized form,
\begin{equation}
\label{GW}
  |\Psi\rangle = \prod_i |\psi_i\rangle, \qquad
|\psi_i\rangle = \sum_{p_1,p_2}
\frac{c_{p_1,p_2}^{(i)}}{\sqrt{p_1! p_2!}} \left(a_{1,i}^\dag\right)^{p_1}
\left(a_{2,i}^\dag\right)^{p_2} | \emptyset \rangle
\end{equation}
where $| \emptyset \rangle$ denotes the vacuum state, $a_{f,i} |
\emptyset \rangle =0$ for any $i$. A time-dependent variational
procedure \cite{Buonsante_JPA_41_175301} shows that the complex
coefficients $c_{p_1,p_2}^{(j)}$ obey a semiclassical dynamics of
the form \cite{Damski_PRL_90_110401}
\begin{eqnarray}
i \dot c_{p_1,p_2}^{(j)} &=&  \left[\frac{U_1}{2} p_1 (p_1-1) + \frac{U_2}{2} p_2 (p_2-1) + U_{1,2} p_1 p_2 
\right]  c_{p_1,p_2}^{(j)} 
-J_1\left[\sqrt{p_1+1}\, c_{p_1+1,p_2}^{(j)} \bar \Phi_1^{(j)} \right.
\nonumber\\
&+& \left . \sqrt{p_1}\, c_{p_1-1,p_2}^{(j)}
\Phi_1^{(j)}\right]-J_2\left[\sqrt{p_2+1}\, c_{p_1,p_2+1}^{(j)} \bar
\Phi_2^{(j)} +\sqrt{p_2}\, c_{p_1,p_2-1}^{(j)}
\Phi_2^{(j)}\right] \label{dG}\\
\qquad \Phi_1^{(j)}&=&\sum_{\ell \sim j} \sum_{p_1,p_2} \sqrt{p_1}
\, \bar c_{p_1,p_2}^{(\ell)} c_{p_1+1,p_2}^{(\ell)} \qquad
\Phi_2^{(j)}=\sum_{\ell \sim j} \sum_{p_1,p_2} \sqrt{p_2} \, \bar
c_{p_1,p_2}^{(\ell)} c_{p_1,p_2+1}^{(\ell)} \label{sh}
\end{eqnarray}
where the symbols $\bar \cdot$ and $\sim$ denote complex conjugation
and nearest-neighborhood, respectively, and the quantities in Eq.
(\ref{sh}) are shorthand notations to simplify Eq. (\ref{dG}).

The ground-state of the system is the lowest-energy normal mode of
Eqs. (\ref{dG}), which can be found as the ground state of the
mean-field Hamiltonian
\begin{eqnarray}
\label{Hmf} {\cal H} &=&  \sum_{i,f} \left\{\frac{U_f}{2} n_{f,i}
(n_{f,i}-1-\mu_f)   -  J_f \left[
a_{f,i}^\dag (\alpha_{f,i+1}+\alpha_{f,i-1}) + a_{f,i}
(\bar \alpha_{f,i+1}+\bar \alpha_{f,i-1}) \right]\right\}\nonumber\\
&+& U_{1\,2} \sum_i  n_{1,i}\, n_{2,i}
\end{eqnarray}
subject to the self consistency constraint $\alpha_{f,j} =
\langle\Psi|a_{f,j}|\Psi\rangle=
\langle\psi_j|a_{f,j}|\psi_j\rangle$. Note indeed that the
eigenstates of Hamiltonian (\ref{Hmf}) have the form in Eq.
(\ref{GW}). A few points are worth observing here. First of all, the
explicit presence of the (species-specific) chemical potentials
$\mu_f$, which act as Lagrange multipliers fixing the total
populations $N_f$. This is explicitly needed since Hamiltonian
(\ref{Hmf}) does not commute with the total populations $N_f =
\sum_j n_{f,j}$, unlike Hamiltonian (\ref{Hamiltonian}). Second, the
mean-field theory can be equivalently obtained from Eq.
(\ref{Hamiltonian}) by means of the so-called {\it decoupling
approximation}, consisting in the substitution $a_{f,i}^\dag a_{f,j}
\longrightarrow a_{f,i}^\dag \alpha_{f,j} + \bar \alpha_{f,i}
a_{f,j}  - \bar \alpha_{f,i}\alpha_{f,j}$. More interestingly, we
note that the above approach treats the on-site terms of Hamiltonian
(\ref{Hamiltonian}) exactly. This in particular means that the
cross-correlation function $C_{1,2}$ introduced in Eq.
(\ref{DeltaN}) below makes sense also for our decoupled trial state
(\ref{GW}). We emphasize the explicit site dependence of our method.
That is, unlike earlier works adopting the Gutzwiller approximation
\cite{Altman_NJP_5_113,Chen_PRA_67_013606,Isacsson_PRB_72_184507},
we do not force translational invariance on Eq. (\ref{GW}), and
hence do not end up with an effective single site theory, which
would be unable to describe the structure of phase-separated or
generically inhomogeneous configurations. Since we are interested in
the effects of an increasing $U_{1,2}$ on a given system, we do not
look for the ground state of Hamiltonian (\ref{Hmf}) at fixed values
of the chemical potentials, as it is usually done. Rather, we adjust
the $\mu_1$ and $\mu_2$ so that we obtain the desired fillings. This
procedure, by the way, avoids the so-called {\it species depletion
problem} affecting single-site theories \cite{Chen_PRA_67_013606,Isacsson_PRB_72_184507}.

The quantities $\alpha_{j,f}$, referred to as {\it local order parameters}, are directly related to several interesting physical quantities, 
such as the one-body density matrix -- related in turn to the condensate fraction of the system and to the interference pattern observed in experiments --
and the superfluid fraction.

In the simple case of a single-species homogeneous lattice one recognizes two phases, corresponding to the local order parameters being all zero or all finite. These situations are readily identified with the insulating and superfluid phases, respectively. The resulting phase diagram turns out to be qualitatively correct, and its quantitative agreement with the exact result improves with increasing dimensionality of the lattice.

Being an approximation, the mean-field approach suffers from some limitations. For instance, site decoupling makes the correlation between two sites of a homogeneous lattice independent of their distance, which results in the superfluid phase being always condensate. This clearly does not apply to 1D lattices, where the power law  decay of the two-site correlations prevents long range order in the thermodynamic limit. Despite this artifact, the mean-field approach proves useful also on one dimensional systems. Indeed, at the finite --- however large ---  experimentally relevant sizes, the mean-field prediction provides an acceptable approximation to the slow decay of the exact correlations. This for instance results in qualitatively correct interference patterns, characterized by a sharp peak at zero momentum \cite{decay}. Such a peak is completely washed out in the Mott-insulating regime.

Furthermore, the local nature of the mean-field order parameters allows for configurations where different portions of the system exhibit different phases,  thus proving effective in capturing the geometric fluctuations that add to the quantum ones in the presence of inhomogeneity. Different phase domains  typically occur in the presence of site-dependent (possibly random) local potentials, but are possible also on homogeneous lattices when the translation invariance is spontaneously broken.

The comparison of the results presented here and in Ref. [27]
with those in Refs. [26,29] is a clear evidence of the validity 
of the qualitative picture provided by the mean-field approach.

For the sake of completeness it should be mentioned that the description of a boson-boson mixture such that both of the atomic species are in a Mott state reveals a further limitation of the mean-field approach.
In this case an unrealistic degeneracy for the ground state of the system is predicted, which can be resolved only taking into account second-order quantum fluctuations [22, 36]. However, the conditions for this artifact to occur are rather specific (the total boson population has to be commensurate with the lattice size) and are never considered in the following.

\section{Results} 
\label{sec.rs}

The zero-temperature phases of the system are determined by the
properties of the ground state of Hamiltonian Eq.
(\ref{Hamiltonian}). We characterize such properties making use of
several quantities. The superfluidity of each species can be
estimated as the stiffness under phase variations
\cite{Roth_PRA_68_023604,Shastry_PRL_65_243,Scalapin0_PRB_47}
\begin{equation}\label{fs}
    F_{\rm s}^{(f)}=\lim_{\theta \rightarrow 0} \frac{1}{t}
    \frac{E(\theta)-E(0)}{N_f\theta^2},
\end{equation}
where $E(\theta)$ is the ground state energy of the Hamiltonian
obtained from Eq. (\ref{Hamiltonian}) by the substitution $a_{f,
j}^\dag a_{f, j+1}+ a_{f, j} a_{f, j+1}^\dag\longrightarrow
e^{i\theta} a_{f, j}^\dag a_{f, j+1} + e^{-i \theta} a_{f, j} a_{f,
j+1}^\dag$ in the hopping term of  species $f$, while  $N_f=\sum_j
\langle n_{f,j} \rangle$ is the population of the same species. The
introduction of the so-called {\it Peierls phases}
\cite{Peierls_ZP_47} is equivalent to the imposition of twisted
boundary conditions.

Two further quantities will be useful in the characterization of the
phases we are going to encounter, namely   the
creation-annihilation cross correlation functions and the local density
fluctuation
\begin{equation}\label{DeltaN}
C_{1,2} =\frac{1}{M} \sum_{i,f} \left(\langle a^\dag_{f,i} a_{\bar f,i}
\rangle-\langle a^\dag_{f,i} \rangle \langle a_{\bar f,i} \rangle\right), 
\qquad
\Delta n_f= \frac{1}{M} \sum_i\left(\langle n_{f,i}
\rangle\right)^2-  \left(\frac{1}{M} \sum_i\langle n_{f,i} \rangle \right)^2.
\end{equation}
where the operation $\bar \cdot$ exchange the two species, i.e. $\bar f=1,2$ 
if $f=2,1$ respectively.
The first quantity is specific to mixtures, accounting for
nontrivial local quantum correlations between particles of different
kind and, as we observed in the previous section,  is well posed for
the Gutzwiller trial state in Eq. (\ref{GW}). The second quantity is
sensitive to  spatial inhomogeneity, and hence it is useful in
the characterization of the quantum emulsion states appearing at
large interspecies interactions. Also, it conveniently signals phase
separation which, in the mean-field approximation, occurs through
the emergence of ground states breaking the translational symmetry
of the Hamiltonian due to nonlinear effects. The expected
homogeneous ground state is recovered as a symmetric superposition
of the set of  degenerate symmetry-breaking states obtained from
each other by a lattice translation.

Finally, the interference pattern in the experimental absorption images 
of each species is basically determined by the Fourier transform of the relevant one-body density matrix  summarizing the first-order correlations \cite{Gerbier_PRA_72_053606},

\begin{equation}
\label{Sk}
S_f({\bf k}) = \frac{1}{M}\sum_{j\,\ell} e^{i {\bf k}\cdot({\bf r}_j-{\bf r}_\ell)}\rho_{j\,\ell}^{(f)}, \qquad \rho_{j\,\ell}^{(f)} = \langle a_{f,i}^\dag a_{f,j}\rangle
\end{equation}
where ${\bf r}_j$ is the spatial position of the $j$-th optical
lattice site, ${\bf k}$ is the (momentum) coordinate in the
absorption image. As we will illustrate shortly, these results provide a reasonable
description of the system, despite the above-mentioned artifact prediction of a finite condensate fraction.

As we mention above, we focus our attention on a combination of
Hamiltonian parameters where several different and interesting
phases take over as the interspecies repulsion $U_{1,2}$ is
increased. As to the remaining parameters, we choose $U_1 = 1$ as
our energy scale, $U_2=\infty$ (hard-core bosons), $J_1=J_2=0.05$,
$N_1/M=1$ and $N_2/M=0.4$. In order to rule out contributions
arising from geometric inhomogeneity we choose a homogeneous lattice
comprising $M$ sites, with periodic boundary conditions.
\begin{figure}
\begin{center}
\resizebox{0.95\columnwidth}{!}{
\includegraphics{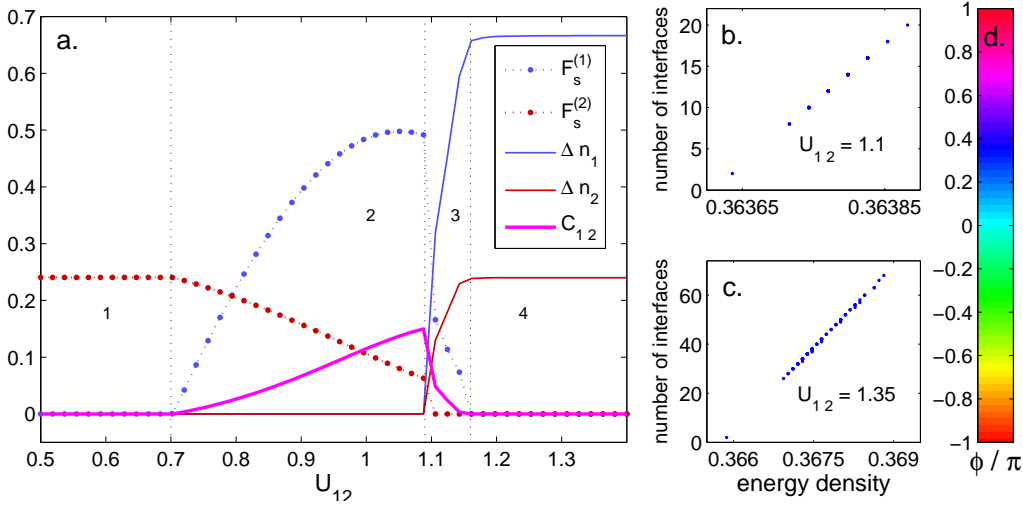} } \caption{a.
Phase diagram of the system as obtained for a lattice with $M=2000$.
The plotted quantities, discussed in Sec. \ref{sec.rs}, allow us to
identify four different quantum phases; Panels b. and c. refer to
regions 3 and 4 of the phase diagram, respectively, and demonstrate
the linear relation between the system energy and the number of
phase interfaces in the quantum emulsion states the system gets
trapped into. The panels show 100 data obtained for a lattice with
600 sites. Panel d. shows the color code for the phase of the local
order parameter characterizing the emulsion droplets in Fig.
\ref{fig:2}. } \label{fig:1}
\end{center}
\end{figure}

The above described quantities allow us to identify 4 different
phases for the system, as it is clear from Fig. \ref{fig:1}. When
$U_{1,2}$ is sufficiently small the situation is similar to the
non-interacting limit $U_{1,2}=0$, in which species 2 is superfluid
and species 1, being strongly interacting and at unitary filling, is
in a Mott-insulating state. Since the two species are virtually
noninteracting, the cross-correlation $C_{1,2}$ vanishes. The
different quantities characterizing the system are independent of
$U_{1,2}$ and each species behaves as if it was alone in the
lattice. In this situation there are no density fluctuations, as it
is expected for single-species systems on homogeneous lattices.

As soon as the interspecies interaction exceeds a first critical
value, $U_{1,2}'\approx 0.7$, it becomes sufficiently strong to
induce a nonzero cross correlation. The interaction with the
hard-core bosons {\it melts} the Mott insulator, and the atoms of
species 1 enter a superfluid phase. This can be intuitively
explained by observing that, owing to the large interspecies
interaction, the simultaneous presence of two atoms of different
species on the same site is energetically unfavourable. Hence the
atoms of species 1 are so-to-say {\it expelled} from the sites
occupied by hard-core bosons, and  give rise to a nonzero superfluid
fraction. Since $U_1$ is still sufficiently larger than $U_{1,2}$,
the bosons of both species remain delocalized and homogeneous, as
demonstrated by the vanishing of $\Delta n_f$. 
It is interesting to observe that
the increase of $F_{\rm s}^{(1)}$ in this phase corresponds to a
decrease of $F_{\rm s}^{(2)}$, as if the hard-core bosons felt the
presence of soft-core bosons like a viscous medium slowing down
their motion through the lattice.

When the interspecies interaction exceeds a second critical value,
$U_{1,2}''\approx 1.09$, the mean-field ground-state of the system
breaks the translational symmetry of Hamiltonian (\ref{Hmf}), which
signals the occurrence of phase-separation. This is recognized by
the nonzero value of the average local density fluctuations of both
species. As we mention above, the expected homogeneous density is
recovered through a symmetrization of the symmetry-breaking state.
As long as the interspecies interaction does not exceed the further
critical value  $U_{1,2}'''\approx 1.16$, the two atomic species
remain correlated, $C_{1,2}>0$, and coherent. However, while the
soft-core bosons are superfluid, the hard-core bosons loose their
superfluidity, $F_{\rm s}^{(2)}=0$. For $U_{1,2} >U_{1,2}'''$ both $F_{\rm s}^{(1)}$ and
$C_{1,2}$ vanish: the system has reached a completely
phase-separated state where no atoms of species 1 are found at sites
hosting hard-core bosons, and {\it vice-versa}.

The above scenario is qualitatively very similar to what is obtained
in Ref. \cite{Hebert_PRA_76_043619}, where a Bose-Fermi mixture is
analyzed by means of  more precise but much more numerically
demanding quantum Monte Carlo simulations. The four phases, with 
a sufficiently large value of $U_2$ instead of an infinitely large one, 
have been proven to be stable under small variations of the interaction parameters.

A very interesting issue recently raised by Roscilde and Cirac
\cite{Roscilde_PRL_98_190402} concerns the possibility that
strongly-interacting mixtures exhibit a  complex low-energy
landscape challenging  very efficient relaxation dynamics such as
that inherent in quantum Monte Carlo simulations. As a result,
usually efficient minimization algorithms fail to converge to the
configuration attaining the minimum energy, and get virtually
trapped into low-energy {\it quantum emulsion} states consisting of
a random arrangement of {\it droplets} characterized by different
phases of the two atomic species. The same phenomenology is captured
by the Gutzwiller mean-field approach, as discussed in Refs.
\cite{Buonsante_PRL_100_240402} and \cite{Menotti_PRL_98_235301} for
strongly interacting bosonic mixtures and single-species bosons with
long range dipolar interactions, respectively.

For interspecies interactions exceeding the phase-separation
threshold, $U_{1,2}''$ the minimization algorithm we adopt gets
trapped in quantum emulsion states, which exhibit different features
in regions 3 and 4 of the phase diagram in Fig. \ref{fig:1}. In
order to illustrate this, we discuss the properties of the quantum
emulsion configurations at two representative values of the
interspecies interaction, $U_{1,2}=1.1$ and $U_{1,2}=1.35$. We first
of all notice that, as discussed in Refs.
\cite{Roscilde_PRL_98_190402,Buonsante_PRL_100_240402} the energy of
quantum emulsion states exhibits a clear linear dependence on the
number of interfaces between neighbouring droplets, as it is shown
by panels b. and c. of Fig. \ref{fig:1}. Clearly, the minimum energy
is attained by a configuration featuring the lowest possible number
of phase interfaces, i.e. two. We emphasize that the lowest-energy
states in Fig. \ref{fig:1} b., c. have been obtained by including an
{\it ad hoc} constraint on the number of interfaces. Indeed, the
presence of a large number of low-energy metastable states makes the
reaching of the true ground state a very hard task.

\begin{figure}
\begin{center}
\begin{tabular}{cc}
\resizebox{6.8cm}{!}{ \includegraphics{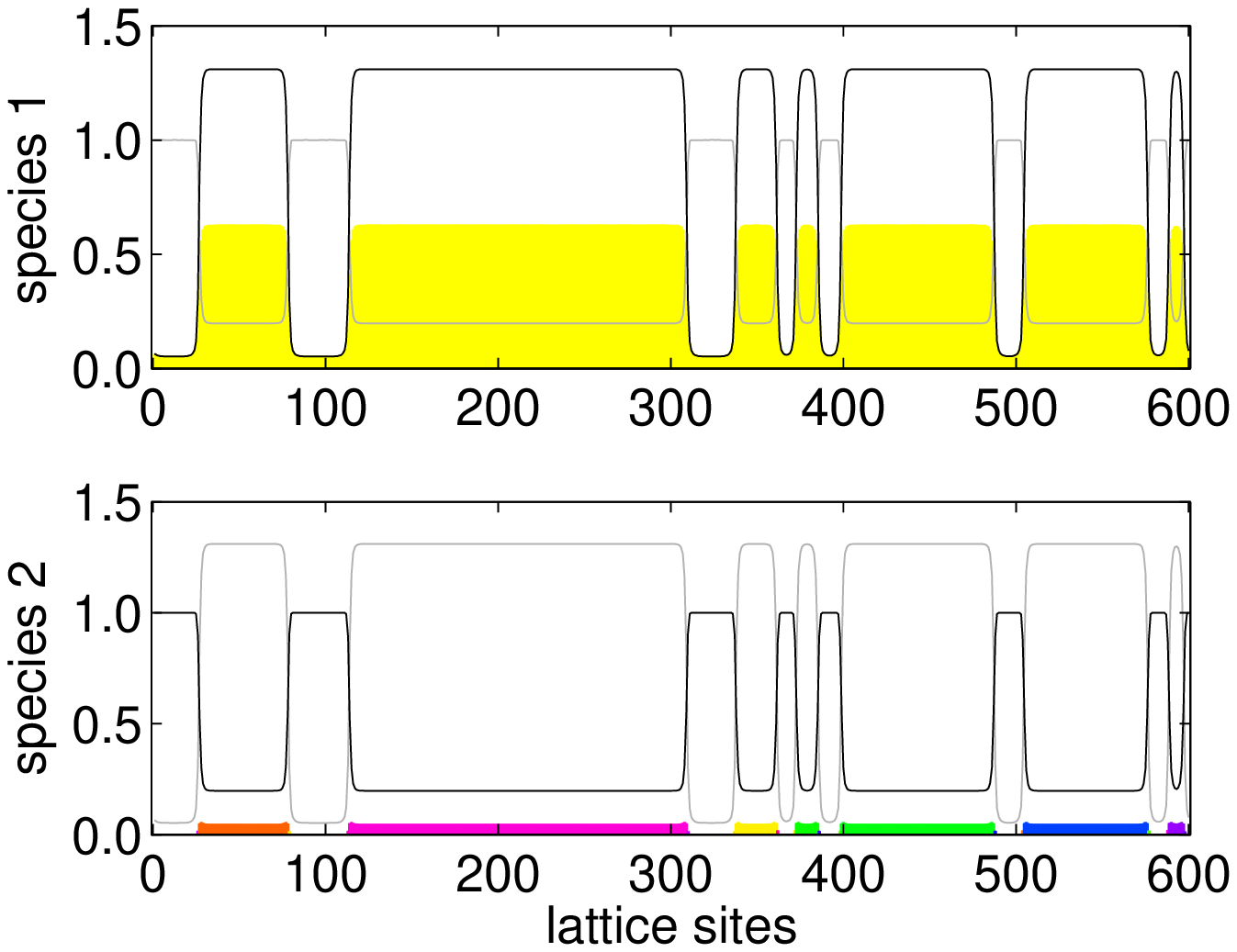} } &
\resizebox{6.8cm}{!}{ \includegraphics{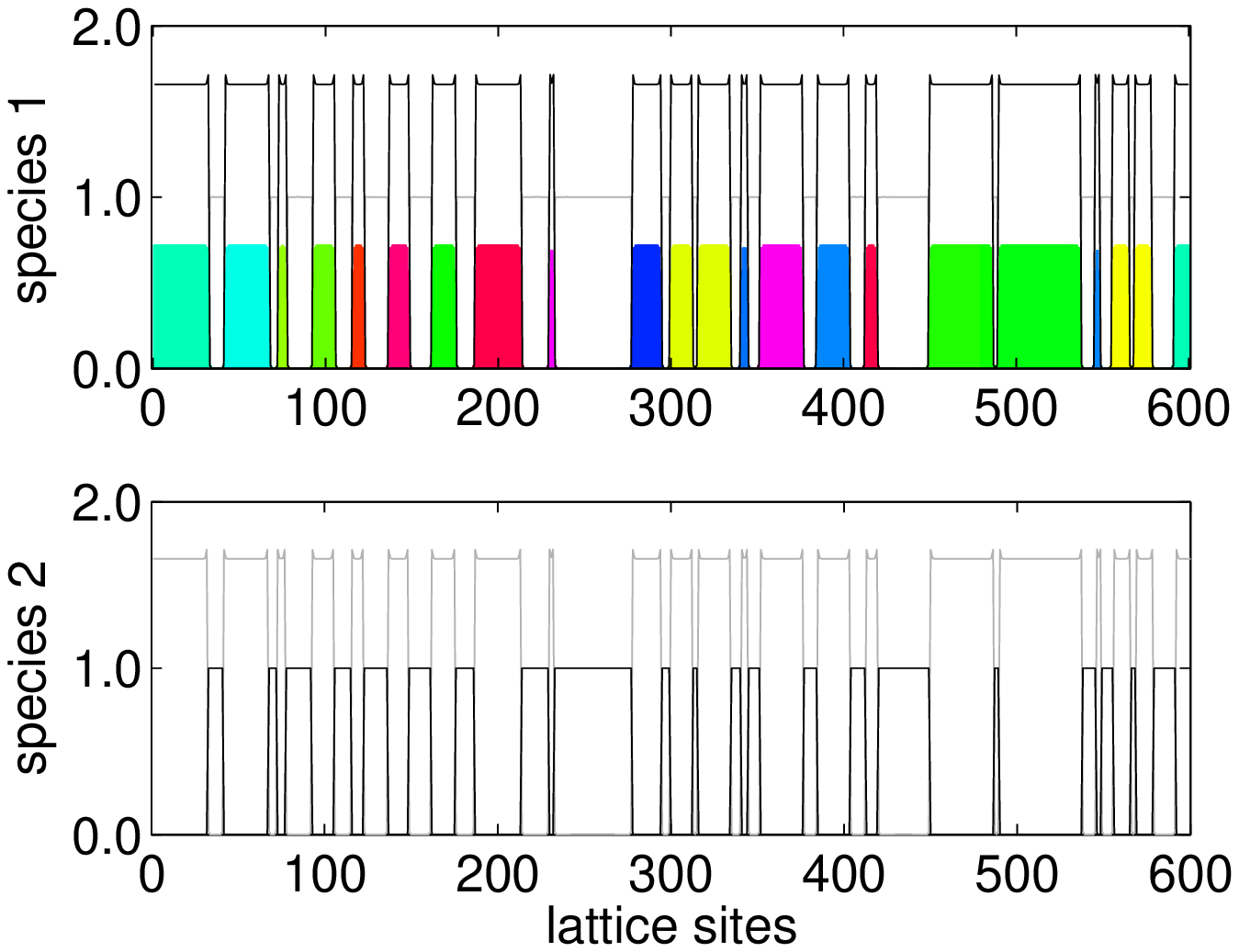} }
\end{tabular}
\caption{Typical quantum emulsion states in regions 3 (left,
$U_{1,2}=1.1$) and 4 (right, $U_{1,2}=1.35$) of the phase diagram in
Fig. \ref{fig:1} a. In  both cases the top and bottom panels focus
on the soft- and hard-core bosons, respectively. The black solid
line represents the local site occupation $\langle n_{f,j}\rangle$,
whereas the coloured areas are related to the complex local order
parameter, $\langle a_{f,j} \rangle=|\alpha_{f,j}|e^{i
\varphi_{f,j}}$. Specifically, the height of the area is
$|\alpha_{f,j}|^2$, while the color is associated to the phase
$\varphi_{f,j}$ through the colorbar in Fig. \ref{fig:1} d. }
\label{fig:2}
\end{center}
\end{figure}

Fig. \ref{fig:2} shows the appearance of the typical quantum
emulsion state at $U_{1,2}=1.1$ (left) and $U_{1,2}=1.35$ (right).
In both cases the upper and lower panels focus on soft- and
hard-core bosons, respectively. The solid black line corresponds to
the local density $\langle n_{f,j}\rangle$, whereas the coloured
areas represent the complex local order parameter
$\alpha_{f,j}=|\alpha_{f,j}|e^{i \varphi_{f,j}}=\langle
a_{f,j}\rangle$. More precisely, the height of such areas is
determined by $|\alpha_{f,j}|^2$ whereas the colors denote the phase
$\varphi_{f,j}$ according to the colorbar in Fig. \ref{fig:1}d. 
We note that each droplet can be characterized by an unique
value of the phase, which can vary arbitrarily from droplet
to droplet.
A light gray solid line representing the local density of the species
analyzed in detail in the other panel is also drawn for comparison
in each panel. In both cases one can recognize the presence of
``droplets'' of two kinds.

For $U_{1,2}=1.1$  the hard-core species comprises irregularly 
interleaved superfluid and insulating domains, which prevents a
superfluid flow.  Indeed, on a one dimensional system such as the 
one under investigation, $\alpha_{2,j}$ must be finite at every site 
in order to produce a nonzero $F^{(2)}_{\rm s}$ \cite{Buonsante_PRA_76_011602}.
Hence the hard-core bosons are in a kind of {\it glassy insulator}, as we discuss in more detail below.
Conversely, the phase of soft-core bosons could be defined as a {\it disordered superfluid}. Indeed it also consists of irregularly interleaved domains, which however always exhibit a superfluid character since 
the value of $\alpha_{i,2}$ never vanishes.

For $U_{1,2}=1.35$ the two atomic species never occupy the same
lattice site, i.e. they are completely phase separated. Mott-like
hard-core droplets are randomly intermingled with superfluid soft-core
domains. Hence in this case the soft-core species is in a
Bose-glass-like phase.

In both  the examined cases the droplets composing the quantum
emulsion are arranged in a random fashion. This feature is quite
interesting, and demonstrates the {\it spontaneously disordered}
nature of the low energy configurations in strongly interacting
atomic mixtures \cite{Roscilde_PRL_98_190402}. Recall indeed that the original Hamiltonians do not
contain any explicit disorder source. Even ignoring the
configurations which are equivalent through a lattice translation, a
virtually infinite number of spatial  arrangements of droplets are
possible with essentially the same energy, very close to the ground
state. Furthermore, in cases where $\alpha_{f,j}$ is finite only
inside one of the two classes of droplets --- as for species 2 at
$U_{1,2}=1.1$ and for species 1 at $U_{1,2}=1.35$ ---, a further
source of energy degeneracy lies in the phase $\varphi_{f,j}$ of the
order parameter. Note indeed that in the above mentioned cases such
phase has a constant value within each droplet, but varies across
different droplets. This is explained by observing that  each
droplet can be well approximated as a virtually isolated and almost
uniform system, owing to the vanishing of the order parameter in the
surrounding droplets. 
In this situation one expects the chosen
periodic boundary conditions to be irrelevant.
In a uniform system an overall constant phase has no effect 
on the observables, and in particular
on the system energy. In the present case different droplets have in
general a different phase, which manifests itself in observables
involving different lattice sites, e.g. like 
\begin{equation}
\label{corrph}
\langle a_{f,j}^\dag
a_{f,\ell}\rangle= |\alpha_{f,j}|\,| \alpha_{f,\ell}| e^{i
(\varphi_{f,\ell}-\varphi_{f,j})}
\end{equation}
 This has virtually no effect on
the energy, since it gets contributions from terms where $\ell$ and
$j$ are nearest neighbours. Indeed when both sites belong to the
same droplet the equal phases  cancel out, while when the sites are
on different sides of a droplet boundary the possible phase
difference is canceled by the virtual vanishing of the order
parameter outside the droplet. 
It is also clear that small variations of the phases $\varphi_{f,j}$
result in an increase of the energy, which demonstrates the
local-minimum character of these configurations \cite{NotePhases}.
 
Interestingly, the different phases of the droplets can play an
important role in the experimental measure of the system coherence.
This is usually obtained by imaging the ultracold atomic cloud after
a few milliseconds of free expansion. 
The interference pattern in the absorption image is substantially described by the Fourier transform of the one-body density matrix summarizing the two-site correlations \cite{Gerbier_PRA_72_053606}, Eq. (\ref{Sk}), whose mean-field form is
\begin{equation}
S_f({\bf k}) =  \frac{1}{M}\sum_j \left(n_{f,j}-|\alpha_{f,j}|^2\right) + \frac{1}{M} \left|\sum_j e^{i ({\bf k}\cdot {\bf r}_j -\varphi_{f,j})}
|\alpha_{f,j}|\right|^2
\end{equation}
We once again recall that in the homogeneous case the mean-field approach results in two-site correlations independent of the site distance, which   leads to an artifact finite condensate fraction in the thermodynamic limit of one dimensional systems. On finite systems this artifact is less serious, since the exact result is anyway very slowly decaying with inter-site distance \cite{decay}. For instance, in both the mean-field and exact cases $S_f({\bf k})$ exhibits a large peak at ${\bf k}=0$ surrounded by small secondary peaks.
Therefore the (incipient) long-range order predicted by the mean-field approach turns out to be a reasonable  approximation to the expected quasi-long-range order on homogeneous systems. 

In the case of single a quantum emulsion state such as those depicted in Fig. \ref{fig:2} 
the correlations in Eq. (\ref{corrph})
would be also finite at arbitrary distance, however modulated by a phase 
factor. This is in contrast to what one would expect in a disordered,
strongly interacting regime. Nevertheless, the expected short range 
correlations are to some extent captured by our mean-field approximation.
Indeed, it seems reasonable to assume that the state of the system consists of a superposition of many almost-degenerate low-energy quantum emulsion states. These will be characterized by different phase-interface arrangements compatible with the typical droplet size dictated by the system parameters, as well as by different phases of the local order parameter within each droplet.
This results in $\rho_{ij}\approx |\alpha_i| |\alpha_j| = |\alpha_i|^2 > 0$ for sites
$i$ and $j$ belonging to the same droplet, and $\rho_{ij} = 0$ otherwise, which
shows that the  correlation lenght is comparable with the average droplet size, as expected.
When this size is of the order of a few lattice sites we expect the absorption image of a quantum emulsion phase to be hardly distinguishable from that of a Mott-insulating phase. This presents some analogies with the case of the {\it Bose-glass} phase induced in a single-species system by (quasi)random local potentials \cite{Lye_PRL_95_070401,Buonsante_LP_18_653}.

\section{Conclusion} \label{sec.end}

In summary, we have discussed the occurrence of four different
phases in a 1D lattice bosonic mixture as a result of the increase
of a single parameter, i.e.  the  interspecies repulsion. The
features of the different phases have been analyzed by employing
suitable observables, and special attention has been devoted to the
quantum emulsion metastable states occurring at large value of the
interspecies repulsion
\cite{Roscilde_PRL_98_190402,Buonsante_PRL_100_240402}.  Also, we
discussed experimental issues such as the visibility of these
quantum emulsion states, which could greatly reduced by the
randomness of the phase of the order parameter within each isolated
droplet. Owing to the large parameter space, we expect that further
quantum phases can crop out in different regimes. For example, we
verified that for a sufficiently large increase in the hopping amplitudes 
phase 3 is replaced by a different quantum emulsion where both species
behave as a disordered fluid.
Such complex phases and the effect of the dimensionality on phase
segregation will be the subject of future work.

\section{Acknowledgments}
The authors acknowledge fruitful discussions with Tommaso Roscilde.


\bibliographystyle{epj}


\begin{thebibliography}{99}
\bibitem{NotePhases} Rigorously speaking, aligning all the phases of the local order parameters while leaving their moduli unchanged attains the minimum energy for a given droplet arrangement. This shows that states where different droplets are characterized by  different phases cannot be the absolute minumum.
On the other hand, as we discussed above, a  generic  quantum emulsion state 
cannot be the absolute minimum either, because of its many phase interfaces.
\end{thebibliography}
\end{document}